# Reconstructing ecological community dynamics from limited observations

Chandler Ross[1], Ville Laitinen[1], Moein Khalighi[1], Jarkko Salojärvi[2,3,4], Willem M. de Vos[5], Guilhem Sommeria-Klein[1,6,*], and Leo Lahti[1,*]

[1]Department of Computing, University of Turku; 20014 Turku, Finland
[2]School of Biological Sciences, Nanyang Technological University, 50 Nanyang Avenue, Singapore, Singapore
[3]Organismal and Evolutionary Biology Research Programme, University of Helsinki; 00100 Helsinki, Finland
[4]Singapore Centre for Environmental Life Sciences Engineering, Nanyang Technological University, Singapore, Singapore
[5]Faculty of Medicine, University of Helsinki; 00100 Helsinki, Finland
[6]Inria, Univ. Bordeaux, INRAE; F-33400 Talence, France
[*]Equal contribution

## Abstract

Ecosystems tend to fluctuate around stable equilibria in response to internal dynamics and environmental factors. Occasionally, they enter an unstable tipping region and collapse into an alternative stable state. Our understanding of how ecological communities vary over time and respond to perturbations depends on our ability to quantify and predict these dynamics. However, the scarcity of long, dense time series data poses a severe bottleneck for characterising community dynamics using existing methods. We overcome this limitation by combining information across multiple short time series using Bayesian inference. By decomposing dynamics into deterministic and stochastic components using Gaussian process priors, we predict stable and tipping regions along the community landscape and quantify resilience while addressing uncertainty. After validation with simulated and real ecological time series, we use the model to question common assumptions underlying classical potential analysis and re-evaluate the stability of previously proposed "tipping element" in the human gut microbiota.

# 1 INTRODUCTION

Ecological communities are subject to stochastic dispersal, birth, and death processes (Harris et al., 2015), and incessant external perturbations caused by environmental fluctuations. Moreover, interactions between community members (Gonze et al., 2017) or ecological memory (Khalighi et al., 2022) can generate complex dynamics (Fujita et al., 2023; Gonze et al., 2018; Konopka, 2009), such as sudden shifts between alternative stable states characterised by different community compositions. Such multistable dynamics have been reported in diverse ecosystems ranging from freshwater plankton communities (Scheffer, 2020) to human gut microbiota (Hartman et al., 2009; Lahti et al., 2014; Sommer et al., 2017). Attempts have been made to disentangle these different ecological processes from time series (Faust et al., 2018). However, the dynamics of ecological communities are notoriously difficult to model as soon as the community size exceeds a few species. This is especially true of microbial communities, due to the diversity and lability of interspecific interactions (Konopka, 2009). Thus, mechanistic models often fail to predict state shifts or to quantify key dynamical properties such as stability and resilience (Bestelmeyer et al., 2011; Faust et al., 2015).

An alternative approach is to characterise the system's fluctuations statistically while making minimal assumptions about the underlying processes (Arani et al., 2021). This has been used to detect early warnings from time series (Laitinen et al., 2021; Scheffer et al., 2009) and to understand how systems change over time (Raulo et al., 2023). A popular approach is to assume that the dynamics are well characterised by a stationary probability density. This approach has been applied to various complex systems such as climate data (Garcia et al., 2017; Livina et al., 2010) and more recently microbial communities (Costea et al., 2018; Shetty et al., 2017). However, an important shortcoming in these approaches is that they usually conflate the number of modes in the probability density with the number of stable states; they do not account for the possibility that the intensity of fluctuations may vary with the system's state (M. S. Arani et al., 2024). Even more crucially, the application of such models in ecology has thus far either relied on long and dense time series to characterise the dynamics, or made the simplifying assumption that it could be recovered from the stationary density without access to time series data (Costea et al., 2018; Lahti et al., 2014; Livina et al., 2010; Shetty et al., 2017). Yet, even though the availability of time series data is quickly increasing in ecology, the vast majority remain short and sparse due to practical or ethical constraints in sample collection - often no more than a few time points per sampling unit (Lahti et al., 2014; Moreno-Indias et al., 2021).

In order to overcome these limitations, we propose a flexible probabilistic model to reconstruct complex system dynamics from multiple short time series. We show how the model can be used to detect a system's stable and *tipping* regions, quantify multistability and resilience, and address uncertainty. We validate the model with simulated and real data. Finally, we highlight shortcomings in the currently popular techniques for the analysis of community stability and we use our approach to re-examine the stability properties of microbial taxa previously proposed to exhibit bistability: the so-called tipping elements of the human gut microbiota (Lahti et al., 2014).

# 2 | METHODS

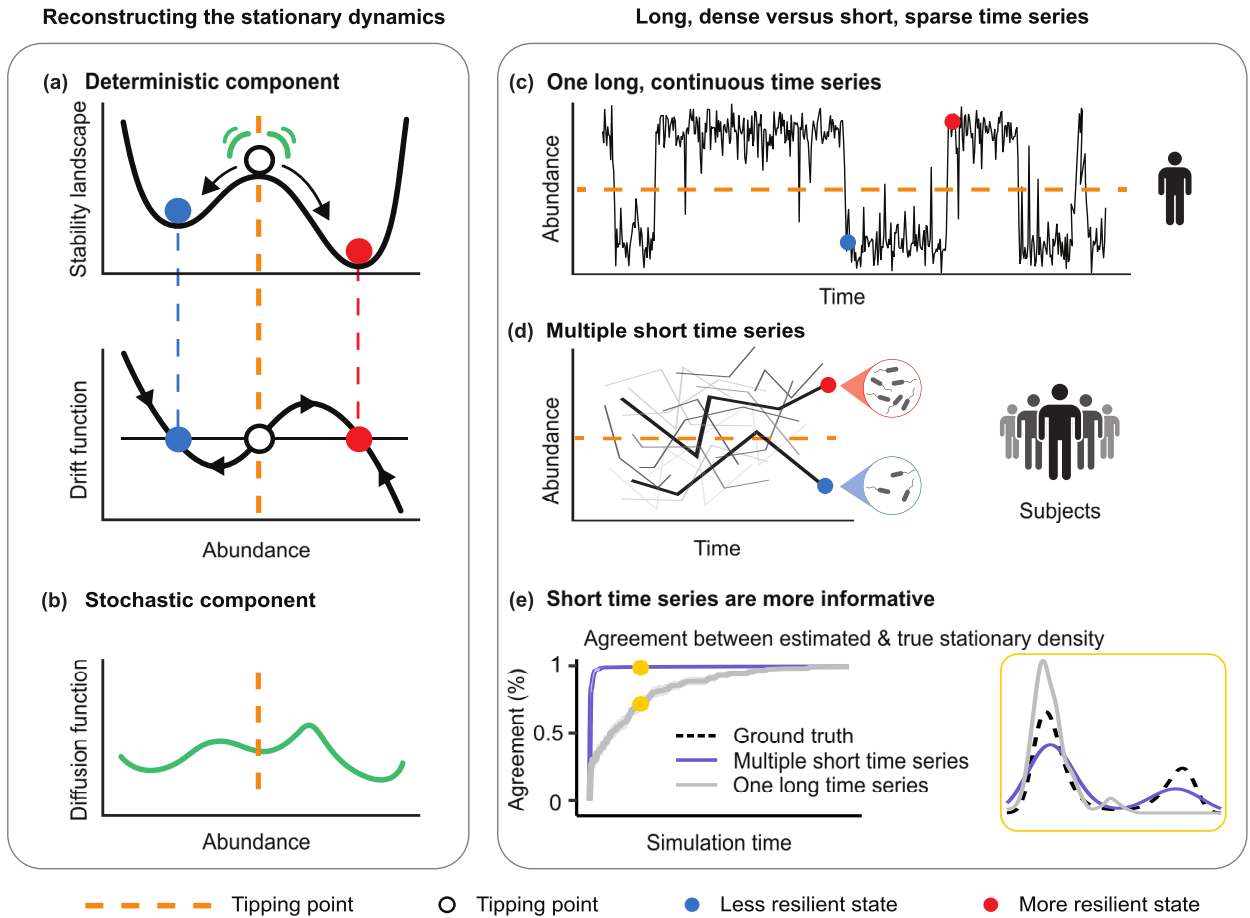

**FIGURE 1** The stability landscape and key concepts. (a) A hypothetical bistable system with two stable states of low and high abundance, separated by an intermediate tipping point (dashed orange line). The dynamics can be described as movement along the stability landscape (upper plot) where at any given time, the system is likely to be found in either the less resilient state (blue) or the more resilient state (red, with a deeper valley) and less likely to be found in the unstable, middle tipping region. The drift function (lower plot) is the negative gradient of the stability landscape; its roots provide the location of the stable states and unstable tipping point of the stability landscape. (b) Movement along the landscape is subject to stochastic perturbations whose intensity may vary with abundance and is described by the diffusion function. (c) A hypothetical long and continuous series of observation from the system defined in (a,b) as assumed by most existing methods for dynamics characterization. (d) A collection of short time series with the same underlying dynamics as in (c) more typical of the data encountered, for instance, in human gut microbiota research. (e) Short, independently sampled time series cover the range of possible measurements in a system more efficiently than a long, continuous time series of the same total length. Both curves show the average agreement with the ground truth stationary density over 50 independent simulations (see *Methods*). Two example densities calculated from the points highlighted in yellow are shown (in blue and grey) compared to the true stationary density of the process from which they were simulated (dashed black line, see also Video V1).

## 2.1 Modelling approach

Our approach is based on the classical representation of ecological dynamics as movement along a stability landscape, also called potential landscape, caused by stochastic fluctuations (Fig. 1a, b) (Arani et al., 2021; Walker et al., 2004). The landscape is a function of the system state, quantified by an indicator variable such as the abundance of a key species, or a higher-level aggregate such as diversity or biomass. Univariate summaries are useful because stability landscapes are generally not well defined in multivariate systems (Rodriguez-Sanchez et al., 2020). Local minima represent stable equilibria, or "stable states". Local maxima represent unstable "tipping points", which mark the boundary between alternative basins of attraction (note that "tipping point" may also refer to a change in the landscape itself in other fields (Van Nes et al., 2016). A lake may, for instance, exhibit distinct states of low and high cyanobacteria abundance, and fluctuate around one of the stable points until a large enough perturbation pushes the system across the tipping point, triggering a transition to the other state (Fig. 1a,b) (Carpenter et al., 2020).

The stability landscape representation assumes that all observations follow the same stationary dynamics. The dynamics can then be decomposed into a deterministic and a stochastic component, using a stochastic process model in which they are quantified by the drift and diffusion functions, respectively (Fig. 1a, b) (Sura and Barsugli, 2002). Drift is the negative derivative of the stability landscape and encodes the shape of the landscape (Fig. 1a). The diffusion encodes the intensity of stochastic fluctuations as a function of system state (Fig. 1b). Classical approaches to inferring this decomposition assume a specific shape for the drift and diffusion. This entails strong assumptions on the landscape shape, the number of possible stable states and fluctuation intensity along the landscape (Livina et al., 2010). Alternatively, non-parametric models allow for a more flexible characterization. However, existing inference methods for these models rely on computing the moments of the trajectory within bins of time points; this relies on long and dense times series that are often not available (Arani et al., 2021; Garcia et al., 2017). In contrast, we propose here a non-parametric approach based on Bayesian inference that allows us to leverage the collections of short time series common in ecological research.

Our approach builds on two key ideas. First, we guide the model with flexible Gaussian Process priors for drift and diffusion (*Methods*), which were recently applied in microbiota research in other contexts (Gerber, 2014; Lloyd-Price et al., 2017). Second, assuming that all samples follow the same stationary dynamics, a collection of short, independently sampled time series can actually cover the state space more efficiently than a single long time series, as consecutive time steps often exhibit strong dependencies (Fig. 1c-e, Video V1). As a result, our approach can use remarkably fewer observations than the commonly used alternative techniques.

## 2.2 Probabilistic model

Our approach employs a drift-diffusion stochastic differential equation model, which assumes that the stationary dynamics can be decomposed into deterministic and stochastic components (Kloeden et al., 1992). We assume an underlying stochastic differential equation for the process, where the dynamics follow a deterministic drift function but are perturbed from this path by a stochastic diffusion function. The mathematical representation of this is given by the Langevin equation (Rinn et al., 2016)

$$\mathrm{d}x(t) = f(x(t))\mathrm{d}t + \sqrt{g(x(t))}\mathrm{d}W(t), \tag{1}$$

where $x(t)$ is the process being modelled at time $t$, $f$ is the drift, $g/2$ is the diffusion, and $W$ is the Wiener process, i.e. white Gaussian noise. Following the Euler-Maruyama scheme, which we know to be convergent (Kloeden et al., 2012), we can discretise equation 1 so that

$$\Delta x \simeq f(x_n)\Delta t + \sqrt{g(x_n)}\Delta W, \tag{2}$$

where $\Delta W = W_{n+1} - W_n$ is normally distributed according to $\mathcal{N}(0, \sqrt{t})$. This step also enforces the assumption of Markovianity in the model. Since the first term is deterministic, we can equally think of $\Delta x$ as being normally distributed:

$$\Delta x \sim \mathcal{N}\left(f(x_n)\Delta t, \sqrt{g(x_n) \cdot \Delta t}\right). \tag{3}$$

Equation 3 acts as the likelihood in our Bayesian framework.

## 2.3 Gaussian processes

Since these data tend to be very noisy, and stochastic differential equations are governed by two functions whose forms are unknown, they pose a problem for parametric models. Gaussian processes act as non-parametric priors in that they do not assume any particular underlying functional form. Instead, they are entirely characterised by their mean and covariance functions, capable of encoding properties such as smoothness and periodicity. They are also powerful in the sense that we can use them for predictive inference: we can query the function's value at points where we do not have observations. We use them in our model to learn the drift and diffusion functions. Their main downfall is that they scale as $\mathcal{O}(n^3)$, where $n$ is the number of observations, and, as a result, they are not ideal for large datasets (more than a few hundred points). They act as priors over functions, and are denoted

$$\phi(\mathbf{x}) \sim \mathcal{GP}(\mathbf{m}(\mathbf{x}), \mathbf{K}(\mathbf{x}, \mathbf{x}')), \tag{4}$$

where $\phi(\mathbf{x}) = (\phi(\mathbf{x}_1), \phi(\mathbf{x}_2), ...\phi(\mathbf{x}_D))$ is the vector of the function $\phi : \mathbb{R}^D \to \mathbb{R}$, $D \in \mathbb{N}$, evaluated at each of the observations $\mathbf{x}_i$ in $\mathbf{x} \in \mathbb{R}^D$, $\mathbf{m}(\mathbf{x})$ denotes the mean function evaluated over all observations, and $\mathbf{K}(\mathbf{x}, \mathbf{x}')$ is the covariance between all possible observation pairs. The process is called Gaussian because it defines a multivariate normal distribution for any set of points $\mathbf{x} \in \mathbb{R}^D$. We set the mean function to zero.

The Gaussian process kernel $\mathbf{K}$ is a positive semi-definite function that holds the majority of our prior beliefs about the function we want to model. We set the kernel for drift prior to be the sum of an exponentiated quadratic kernel and a linear kernel, that is:

$$\mathbf{K}_f(x, x') = \sigma_q^2 \exp\left(-\frac{1}{2l^2}(x-x')^2\right) + \sigma_b^2 + \sigma_l^2(x-c)(x'-c). \tag{5}$$

The $\sigma_q$, $\sigma_b$ and $\sigma_l$ terms in equation 5 are variance parameters, $l$ is the length scale, and $c$ is a centring parameter. The exponentiated quadratic kernel (first term in equation 5) leads to functions with the desirable property of infinite differentiability. The linear kernel (second and third terms in equation 5) will cause the drift, in a region where there is no data point, to tend towards a line with a slope close to what it had in the adjacent region with data. Its inclusion aids in producing the desired end behaviour: instead of going toward zero, we want the drift to approach plus or minus infinity, which is required if we are to obtain stationary dynamics. The diffusion on the other hand only benefits from the properties of the exponentiated quadratic kernel since we cannot say anything a priori about its end behaviour. The one quality we do require that the diffusion have, however, is non-negativity. It is not straightforward to enforce this in the Gaussian process prior itself. In order to constrain the model, we use the following sampling statement for equation 3:

$$\Delta x \sim \mathcal{N}\left(f(x_n)\Delta t, \sqrt{\exp(\hat{g}(x_n)) \cdot \Delta t}\right), \tag{6}$$

where $\exp(\hat{g}(x_n)) = g(x_n)$. In practice, we set a Gaussian process prior on $\hat{g}(x_n)$ rather than $g(x_n)$, and then transform it back to the actual quantity of interest $g(x_n)$.

## 2.4 Model inference

We base model inference on a number of assumptions. We are concerned here with datasets that comprise several short times series that come from different but related sampling units, such as different study participants in the case of human gut microbiota. To pool information between limited data, we assume that they represent different realisations from the same (or similar) underlying dynamics, and model them jointly. We assume that the first two moments of the stochastic differential equation, the drift and the diffusion, are sufficient in describing the dynamics. In addition, we assume that the time series are stationary, Markovian, and that the data takes sufficiently small time steps to minimise the error attributed by the Euler-Maruyama discretisation.

We set standard Inverse-Gamma(2, 2) priors on all variance parameters and Inverse-Gamma(5, 5) priors on length scale parameters. Inverse-gamma distributions work well for these parameters because they heavily suppress values close to zero but have tails that stretch to slightly larger values. In terms of the length scale parameters, this allows for both curvy, high frequency functions, and practically linear, low frequency functions, depending on the data. This behaviour is desirable in both the drift and diffusion functions. If the system is bimodal but not bistable, a flexible diffusion prior is needed. Conversely, if it is bistable, a flexible drift prior is needed. The linear and exponentiated quadratic kernels' variance priors are the same because, in some cases, they are given equal weight. In the case of a bistable drift, the exponentiated quadratic kernel's learned variance should have a greater influence. On the other hand, in the case of a unistable drift, the linear kernel's variance should take the dominant role.

We performed Bayesian inference using the probabilistic programming language Stan, which uses Hamiltonian Markov chain Monte Carlo. We evaluated the convergence of the model using the statistics available in Stan, namely, the Rhat and n_eff values. In all cases, we used 4 chains with 2,000 iterations per chain.

## 2.5 Characteristic time scale

The time series need to be sufficiently dense to satisfy our modelling assumptions. To assess this, we introduce a time scale that can be calculated from any time series data, provided it can be assumed to follow approximately stationary dynamics. We define this time scale as the apparent time needed to traverse the observed range $d$ of the data through stochastic fluctuations. We expressed this condition as $\sqrt{\langle \Delta x^2 / \Delta t \rangle \cdot \Delta t} = d$, where $\langle \Delta x^2 / \Delta t \rangle$ is the average of $(x(t+\Delta t) - x(t))^2 / \Delta t$ over the time series, which yields $t_c = d^2 / \langle \Delta x^2 / \Delta t \rangle$. We expressed all time steps used in this paper as a fraction of this time scale. In the context of our model, $\langle \Delta x^2 / \Delta t \rangle$ can be seen as an approximation for $\langle g(x) \rangle$ since $g(x) = \lim_{\Delta t \to 0} \frac{1}{\Delta t} \mathbb{E}[(x(t + \Delta t) - x(t))^2]$ (Friedrich et al., 2011).

## 2.6 Derived quantities

For our purposes, the quantities of interest that can be derived directly from the drift and diffusion functions are the stationary density, stability landscape, multistability, and exit time.

The stationary density tells us how likely we are to find the system in a given state. Given the drift and diffusion (1), it is possible to derive the stationary distribution via (Iacus and Yoshida, 2018)

$$\pi(x) \propto \frac{1}{g(x)} \exp\left\{ 2 \int_{x_0}^{x} \frac{f(y)}{g(y)} dy \right\}, \qquad (7)$$

where $x_0$ is any point in the state space of $x$ and the constant of proportionality is the normalisation factor. The effective potential equivalently summarises the stationary state and is given by:

$$U_{\text{eff}}(x) = -\log \pi(x). \qquad (8)$$

The stability landscape is related to the drift via the transformation (Iacus, 2010):

$$U(x) = -\int_{x_0}^{x} f(y) dy \qquad (9)$$

and reflects the underlying stability of the system.

The multistability posterior reflects the model's confidence in the number of stable states underpinning the dynamics (Fig. 3d,j and Fig. 4c,d). The minimum number of stable states is one, and the posterior probabilities of the number states supported by the model sum to 100 %. When calculating this quantity, we looked at the roots of the posterior drift draws with positive slopes (tipping points) and negative slopes (stable points). In order to obtain a stationary solution, we required that the number of stable equilibria be one more than the number of tipping points in the system. Any draws that did not meet this requirement were excluded from the calculation since classifying their stability would be ambiguous. For each of the remaining draws, we could calculate the posterior of the multistability of the system by counting the number of roots with a negative slope.

Finally, the mean exit time can be determined by solving the second order ordinary differential equation:

$$f(x)\frac{\partial T(x)}{\partial x} + \frac{g(x)}{2}\frac{\partial^2 T(x)}{\partial x^2} = -1, \tag{10}$$

where $T(x)$ is the exit time (Arani et al., 2021). In order to solve (10), we need to impose three boundary conditions: the slope must tend to zero as the state variable approaches $\pm\infty$ and the exit time must be zero at the tipping point. We achieved this using a custom finite difference method to numerically approximate the solution.

## 2.7 Exit time uncertainty estimation

After running the model and obtaining posteriors for the drift and diffusion functions, we can calculate the exit time posterior by computing (10) over each of the posterior draws. The resulting exit time posterior draws can have vastly different tipping point locations which makes it hard to visualise. In order to quantify the uncertainty, we pruned the posterior draws to retain only those draws that had the same tipping point (up to the discretisation step) as the posterior means of the drift and diffusion. Then, starting from the draws with the smallest exit time values and working up, we calculate the lower 60% and 40% credible intervals. In this way, we avoid the draws with extremely large values which would be captured with standard credible intervals and we obtain good lower bound estimates on the exit time for a given tipping point.

## 2.8 Simulation model

The model was tested on the well-studied stochastic process called the cusp catastrophe model. All of its dynamical properties such as its drift, diffusion, stationary density, and so on have analytical forms for comparison (Iacus, 2010). Not only do we know the ground truth, but the model allowed us to test a wide range of drift topologies including those that lead to unistable, bistable, and skewed stationary densities. Those that we expect to find in real data are not limited to any specific case. In particular, it is a stochastic process of the form

$$\mathrm{d}x(t) = r(\alpha + \beta(x(t) - \lambda) - (x(t) - \lambda)^3)\mathrm{d}t + \sqrt{\epsilon}\mathrm{d}W(t), \tag{11}$$

where $\alpha, \beta, \lambda \in \mathbb{R}$, and $r, \epsilon > 0$ are the free cusp parameters.

Short time series are generated by drawing a starting value from the stationary density of the cusp model which is analytically known using inverse transform sampling. We then evolve the system using equation 11 and the Euler-Maruyama approximation scheme with time steps of 0.01 in order to obtain a convergent simulation. These high resolution time series are then subsetted to the desired time step that is larger than the 0.01 resolution. This process was repeated for each short time series.

For the bimodal, unistable case, we used custom drift and diffusion functions and evolved them according to the Euler-Maruyama scheme, sampling the starting points from the approximate stationary density in the same fashion as before. We chose the drift and diffusion as:

$$f(x) = \begin{cases} \exp(-0.08x) - 0.95 & x \leq 0 \\ -0.5x^2 + 0.05 & x > 0 \end{cases} \quad \text{and} \quad g(x) = 0.844 \cdot \exp[-(2x - 0.6)^2]. \qquad (12)$$

## 2.9 Short time series

In order to demonstrate the efficacy of short time series in covering the state space of a system, we simulated 50 independent short and long time series from the cusp model (above), each with the same total simulation time of 250. Applying an expanding window approach to these time series starting from a total simulation time of 1 and increasing by increments of 1, we calculated the Kullback-Leibler divergence between the ground truth stationary density of the system and the density approximated from the data histograms. We then normalised the results and subtracted them from 1 to obtain our agreement measure (Fig. 1e, Video V1). In general, the histograms generated from the short time series converged much more rapidly than those of the long time series. In multistable systems, sampling times must be on the order of the characteristic time scale of the system (see above, characteristic time scale) to observe the alternative stable states of the system in continuous time series. Independently sampled short time series, on the other hand, are likely to sample all stable states on any timescale.

## 2.10 Gut microbiota profiling

We retrieved a targeted subset of previously collected human gut microbiota profiling data from the HITChip Atlas database maintained at the Wageningen University and Research Center (Kumaraswamy et al., 2024; Lahti et al., 2014). In summary, DNA from faecal samples was extracted via the repeated bead beating (RBB) method and hybridised on the short oligos on the HITChip microarray (Rajilić-Stojanović et al., 2009), which has been designed to quantify the relative abundance of 130 genus-level groups that collectively capture the majority of the human gut microbiota variation. This array-based profiling technique provides an alternative to the current sequencing-based profiling techniques with high reproducibility (Rajilić-Stojanović et al., 2009). To work with compositional data, we applied the standard centred log-ratio transformation (CLR) (Gloor et al., 2017). We excluded individuals with reported health issues, severe obesity, antibiotic use, and interventions, obtaining data from altogether 128 adults who had data from two or more time points. For the drift-diffusion model (*Methods*), we additionally excluded time steps of more than 45 days, leaving 39 subjects for dynamical analysis (Fig. 4a,b); this corresponds to a time step of $0.02 \cdot t_c$ (Fig. 3f, see "Characteristic time scale" above). We analysed the 63 most prevalent genus-level groups showing at least 20% prevalence.

# 3 RESULTS

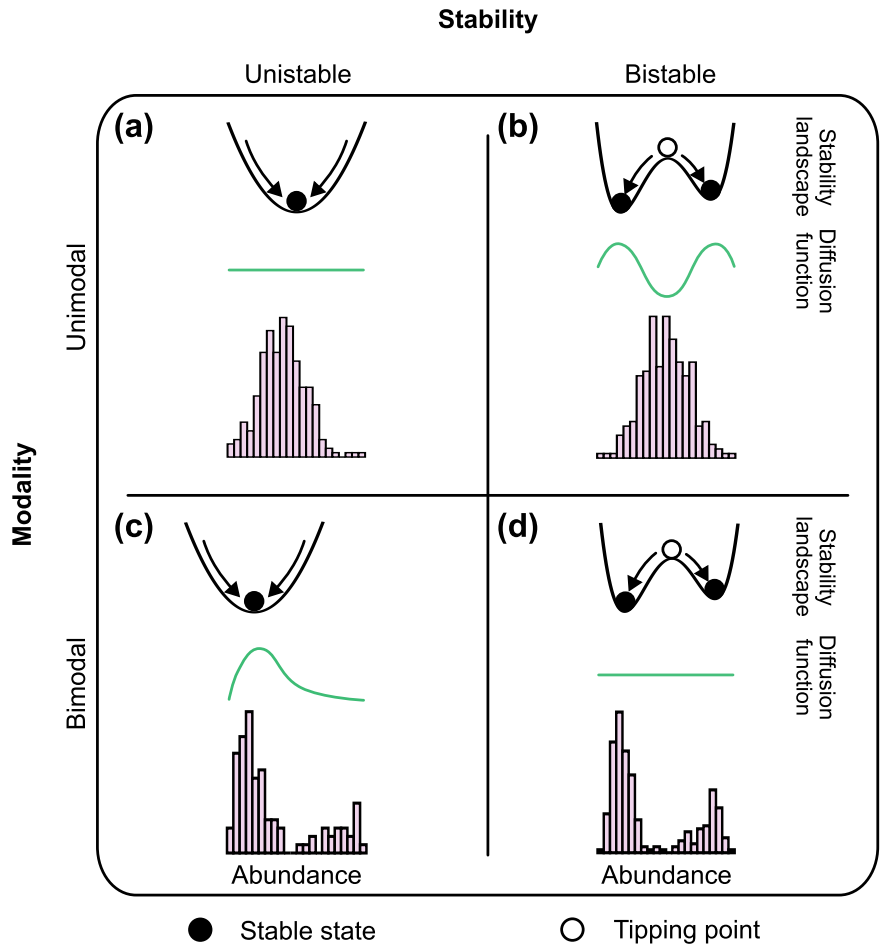

**FIGURE 2** Disentangling modality and stability. (a,d) Usual cases where the bimodality of the data histogram coincides with the bistability of the underlying process: (a) the unistable landscape gives rise to a unimodal histogram and (d) the bistable landscape to a bimodal one. This is the case as long as the diffusion function (green curves) is relatively constant. (b,c) This congruency breaks in the presence of more complicated diffusion functions: with the illustrated diffusion functions, (b) the bistable potential gives rise to a unimodal histogram and (c) the unistable potential to a bimodal one.

## 3.1 Stability and modality

Multistability in ecological systems can arise for instance from interactions between community members (Gonze et al., 2017) or ecological memory (Khalighi et al., 2022). Whereas identifying multistability is important for the prediction and manipulation of ecological systems, this has proven to be challenging in practice. Multistable systems spend most of the time fluctuating around their stable points: this is reflected in the density of observations, which tend to be clustered around stable points. Accordingly, the "effective potential" obtained from the stationary distribution of observations is often used as an approximation for the stability landscape. For instance, the so-called "potential analysis" considers bimodal stationary density as diagnostic for bistable dynamics (Livina et al., 2010). More generally, multimodality in the stationary density has frequently been treated interchangeably with multistability in the ecological literature (Dakos and Kéfi, 2022; Lahti et al., 2014; Livina et al., 2010). However, this assumption can be misleading as, in general, the distribution of observations does not carry enough information to identify multistability. The deterministic and stochastic components of the dynamics may indeed interact to yield non-trivial stationary densities (Fig. 2).

The modes of the observed data can be used to infer stability when the diffusion is approximately constant, that is, when the intensity of fluctuations does not depend on the system state (Fig. 2a,d). However, when stochastic fluctuations are wider close to the stable state than away from it, a unistable system may generate bimodal observations (Fig. 2c). We term this scenario *diffusion-driven* bimodality, in contrast to the more classic *drift-driven* bimodality (Fig. S1). Diffusion-driven bimodality may be expected in ecological systems if, for instance, growth rates and competition increase around the system equilibrium, thus accelerating the dynamics. Increasing fluctuation intensity close to the stable states may similarly conceal a bistable dynamics behind unimodal observations (Fig. 2b).

Hence, the influence of stochasticity cannot be reliably detected by merely examining the static observation frequencies. Access to time series makes it possible to disentangle the stochastic and deterministic components in more general scenarios. Our approach is flexible enough to accommodate many different forms of drift and diffusion, including the skewed and multistable stability landscapes characteristic of microbial communities (Lahti et al., 2014). Although the role of stochasticity on the shape of stationary densities has been widely ignored until recently (Dakos and Kéfi, 2022), the importance of distinguishing between modality and stability in complex systems is being increasingly recognised in the literature (Arani, Carpenter, and van Nes, 2024; Arani, van Nes, et al., 2024; M. S. Arani et al., 2024). We emphasise here non-constant diffusion as a key driver for the incongruency between stability and modality, but this phenomenon could also be attributed to other dynamical properties, such as non-normality or the presence of memory in the distribution of stochastic fluctuations (M. S. Arani et al., 2024). Our approach could possibly be extended to account for these additional factors.

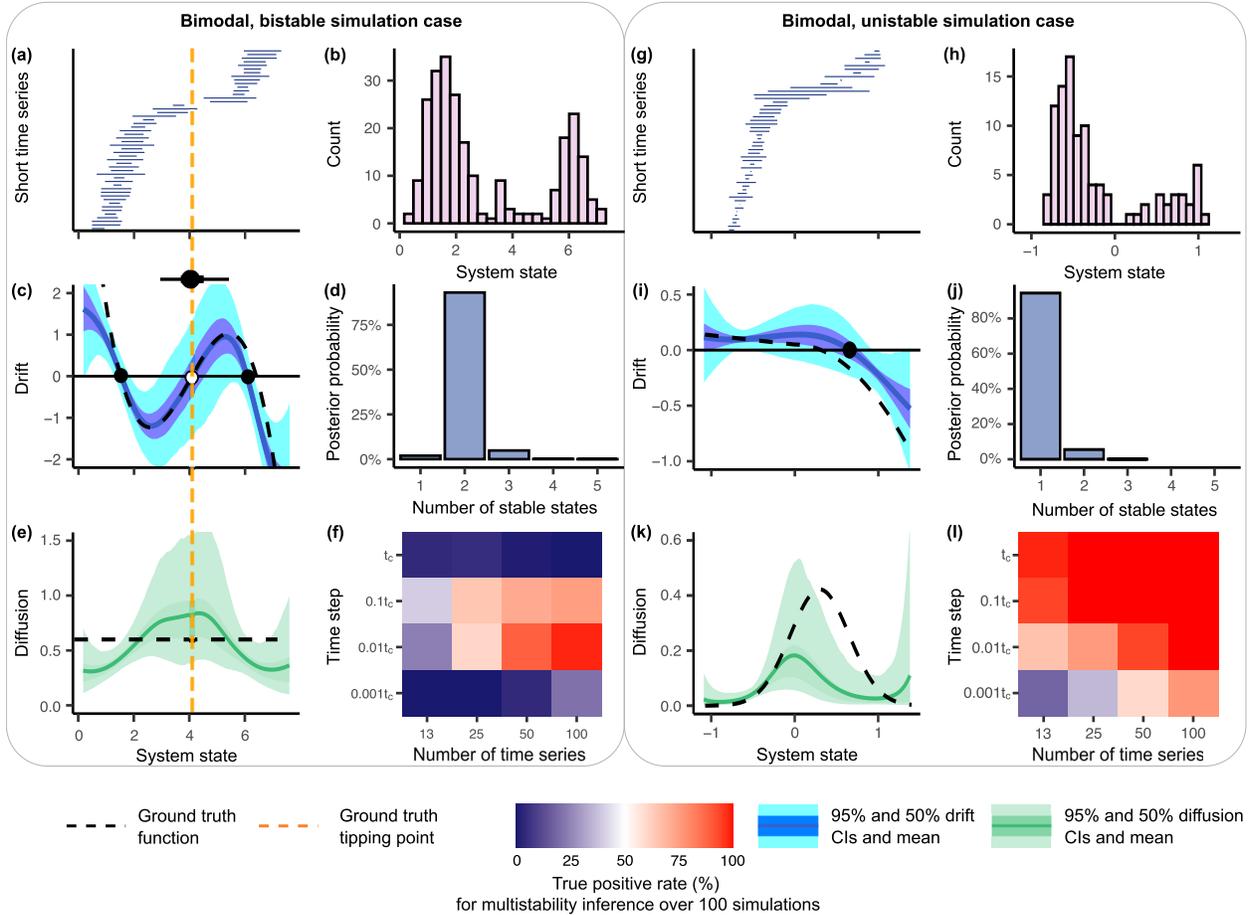

**FIGURE 3** Inferring system dynamics and multistability. (a,g) Short time series (5 points each) simulated from (a) a bistable stochastic process (cusp model), and (g) a unistable process. The known tipping point of the bistable process is represented by a dashed orange line. (b,h) Histograms of the short time series in (a,g). (c,e,i,k) posterior estimates for the (c,i) drift and (e,k) diffusion functions, including the 50% (indigo and red, respectively) and 95% (cyan and salmon, respectively) credible intervals, and the ground truths (dashed black lines). Black and white points on the posterior mean of the drift function indicate stable modes and the tipping point, respectively. The posterior distribution of the tipping point - the tipping region - is summarized by its mean and 95% credible interval in the upper margin of (c). (d,j) Posterior probability of the number of stable states. (f,l) Each cell represents the true positive rate of detecting the stability of the underlying process in 100 independent simulations of 2-point time series, varying the number of time series (i.e., half the number of points) and the sampling time step defined as a fraction of the time scale $t_c$ of the process. Multistability inference is most accurate for intermediate sampling time steps, and generally improves with larger sample size.

## 3.2 Model assessment

We analysed model performance based on the well-established cusp simulation model (Zeeman and Barrett, 1979), which is capable of generating a variety of uni- and bistable dynamics via a third-order polynomial drift and constant diffusion (Fig. 3a-f). We also generated bimodal but unistable dynamics with non-constant diffusion using a custom stochastic differential equation (Fig. 3g-l, equation 12 in *Methods*). In both cases, we simulated short time series and then reconstructed the drift and diffusion functions using our modelling approach. The true drift function and tipping and stable point locations retrieved by our model fell nearly entirely within the 95% posterior credible intervals (Fig. 3c,i). The diffusion generally followed the ground truth but exhibited larger deviations (Fig. 3e,k). More specifically, the diffusion follows the trend already reported in Arani et al. 2021 (Arani et al., 2021): drift-diffusion models tend to underestimate the diffusion in the vicinity of stable points and overestimate it near the tipping point. A low level of fluctuations around stable points can be described either by a strong deterministic component or a weak stochastic one, and distinguishing between these is challenging from sparse data. Similarly, around the tipping point, a high level of fluctuations may be either attributed to a weak drift or a strong diffusion.

The two key factors affecting the performance of the model are the number of short time series and their density, quantified by the time step between two consecutive points (Fig. 3f, l). In general, more and denser time series improved inference. Indeed, shorter time steps are in better agreement with our discretisation approximation (see equation 2 in *Methods*). Nevertheless, inference accuracy dropped again for time steps that were too short to sample the state space efficiently, given the small number of time points per time series we consider (here, only two). We expressed the time step between two points as a fraction of the characteristic time scale of the dynamics $t_c$ (see *Methods*). We found that bistability can be detected with 72% to 89% reliability (proportion of true positive across 100 independent simulations) with as few as 50 time series with two points each for constant simulated diffusion, provided that the time step is on the order of $0.1t_c$ or $0.01t_c$ (Fig. 3f). Bistability inference is therefore robust to sparse data using our approach. Furthermore, if the time step is in the favourable range of about $0.01t_c$ and there are at least 100 short time series, the drift and diffusion estimates are sufficiently accurate to permit the reliable inference of secondary quantities such as stationary densities and stable state resilience, as we describe in the next sections. Finally, inference of unistability is generally robust from short and sparse time series even when the data exhibit bimodal observation densities (Fig. 3l).

## 3.3 Lake Mendota data

To illustrate our approach, we re-analysed the phycocyanin level time series data from Lake Mendota (Wisconsin, USA), a one-year dense time series of measurements made every few minutes over the year 2011 (Carpenter et al., 2020). Phycocyanin level is a proxy for Cyanobacteria abundance and used as a descriptor of the ecological state of the lake. Arani et al. (2021) (Arani et al., 2021) estimated the drift and diffusion for this data set using a non-parametric model, where they calculated the moments of the stochastic differential equation given in equation 1 (see *Methods*, probabilistic model) using a single time series consisting of $6 \times 10^4$ equally-spaced data points (dt ≈ 2.5 mins). In order to compare our approach to theirs, we selected 10 short time series (equally-spaced with dt = 12 days between time series, thus avoiding the non-Markovianity issues raised in Arani et al.) of five equally-spaced time points each (dt = 10 mins between adjacent points). We then showed that the same dynamics can be reconstructed based on these 50 observations extracted from the same data, that is, using three orders of magnitude fewer data points (Fig. S2a,b).

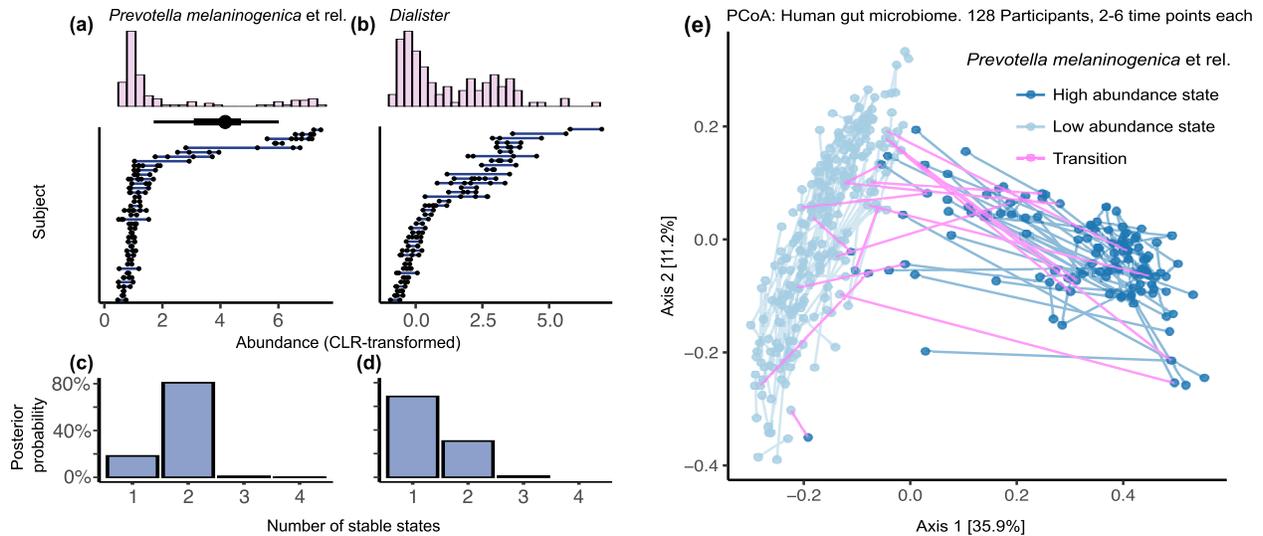

**FIGURE 4** Reanalysis of human gut microbiota stability. (a,b) 39 short and dense time series of the abundance of (a) *Prevotella melaninogenica* et rel. and (b) *Dialister* (each collected from a different subject, with 2-6 time points) on which we ran our model, subsetted from the 128 original time series shown in (e) as described in *Methods*. The empirical histograms are shown in the upper plots, and the range of abundance variation for each subject in the lower plots. In (a), the black horizontal bar summarizes the posterior distribution of the tipping point location identified by the model (most probable location with 50% and 95% credible intervals). (c,d) Posterior probabilities for the number of stable states. (e) Ordination representing the overall community similarity (PCoA, Bray-Curtis index) between 358 human gut microbiota samples from 128 time series, including the subset of 39 dense time series shown in (a,b). Dark and light blue points indicate the high and low abundance stable states of *Prevotella melaninogenica* et rel. identified by our model, respectively. The arrows represent the paths of individual subjects through time. Pink lines indicate transitions across the tipping point identified by our model.

## 3.4 Tipping elements

Recent population studies characterised the landscape of cross-sectional variation in gut microbiota composition (i.e., variation between individuals; Costea et al., 2018). Characterising the stability landscape has been more challenging due to the scarcity of population-level time series data and the lack of suitable methods applicable to limited time series. Our model is addressing this gap. Lahti et al. (2014) have previously reported evidence for the bistability of certain sub-communities of the human gut microbiota, the so-called *tipping elements* (Lahti et al., 2014). This analysis was based on taxonomic profiling of 1,006 western adults, of which 78 had multiple time points (Lahti et al., 2014). We extended this dataset by retrieving additional data for altogether 128 adults with short time series of 2-6 time points each (see *Methods* (Kumaraswamy et al., 2024; Lahti et al., 2014)). In order to respect the Euler-Maruyama approximation used in our approach, we included in the final model only those 39 time series that had sufficiently small time steps (lower than $0.02 t_c$, Fig. 4a,b, *Methods*).

Our drift-diffusion model identified bistability in 5 genus-level groups (out of the 63 groups with highest prevalence; see *Methods*): *Bifidobacterium*, *Eubacterium biforme*, *Prevotella oralis* et rel., *Prevotella melaninogenica* et rel., and Uncultured *Mollicutes* (Fig. 4a,c and Fig. S3). The two stable states of *Prevotella melaninogenica* appear to be strong ecosystem-level drivers as their positions are clearly visible in the landscape of community-level variation across individuals (first two PCoA axes, Fig. 4e). The previously reported tipping elements *Bacteroides fragilis* et rel., *Dialister*, and *Uncultured Clostridiales* were deemed unstable by our model, despite bimodal abundances and in contrast to the evidence for bistability originally reported using a cruder approach on a subset of the same data (Lahti et al., 2014) (Fig. 4b,d and Fig. S4). Moreover, our probabilistic model quantifies uncertainty, which is essential in natural systems where stochasticity and measurement errors make it virtually impossible to determine the number of stable points with full confidence, or determine exact tipping point locations. In such cases, these uncertainties should be presented as an integral part of the results. In particular, the *tipping region*, the interval with an elevated probability for state shifts, may be more useful to report than a tipping point.

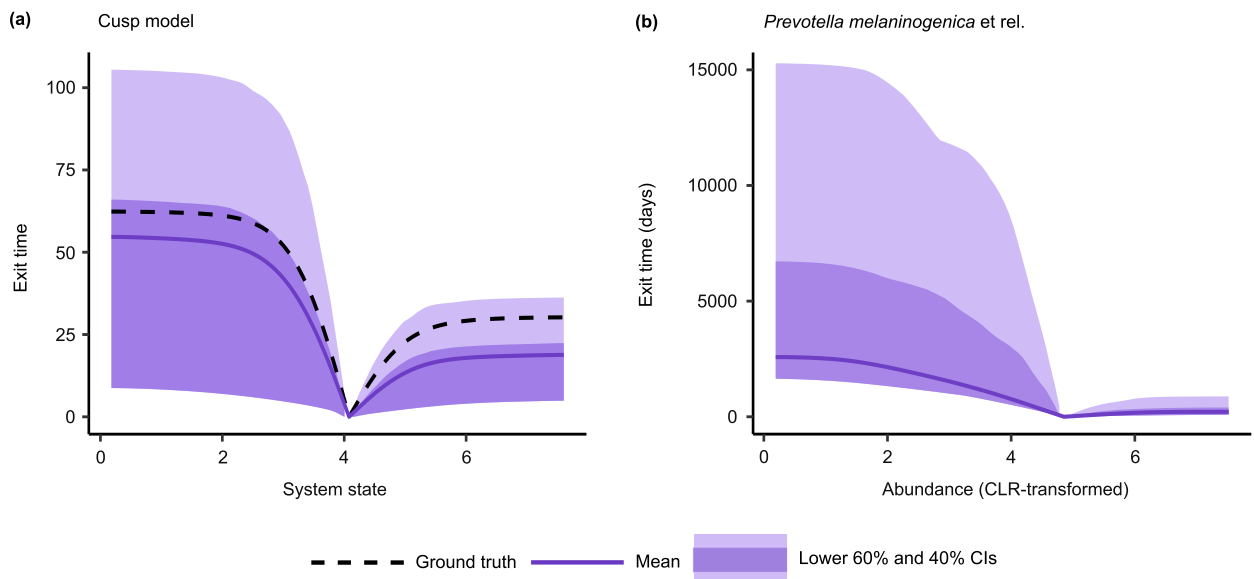

**FIGURE 5** Predicted exit time. (a) Exit times predicted as a function of abundance, calculated from the simulated short time series shown in Fig. 3a. The dashed black line represents the known ground truth. (b) Exit times predicted for the *Prevotella melaninogenica* et rel. data of Fig. 4a. In (a) and (b), the posterior distributions are computed conditionally on the the location of the tipping point, and are represented by their mean (dark purple line) and their 40% and 60% lower credible intervals (dark and light purple regions, which are bounded by the posterior draw with the smallest exit time value).

### 3.5 Resilience and exit times

Resilience quantifies the system's tendency to remain in its current state in the face of perturbations (Nolting and Abbott, 2016). Recently, Arani et al. (2021) (Arani et al., 2021) proposed the expected exit time as a new measure of resilience under continuous stochastic perturbations. This metric quantifies how long the system is expected to remain in its current stable state given its observed fluctuation regime (Scheffer et al., 2015). Conveniently, the exit time can be directly derived from the inferred drift and diffusion functions (Scheffer et al., 2015). The explicit incorporation of stochastic variation sets this measure apart from other measures of resilience.

However, the original formulation relied on long and dense time series and the exit time itself is sensitive to small differences in the drift, diffusion, and tipping point estimates. Here, we provide a probabilistic estimate of exit time, which accounts for the uncertainty arising from limited observations (see *Methods*). Our approach could reliably estimate the exit time as a function of system state from the same simulated short time series as in Fig. 3a (Fig. 5a). Furthermore, the estimated exit times for the bistable *Prevotella melaninogenica* et rel. are in line with the time scales observed in the data (Fig. 5b, Fig. S5). However, the calculation of the exit time and its uncertainty are sensitive to the ratio between drift and diffusion and to the tipping point location. When the diffusion is low compared to the drift, the dynamics become strongly deterministic and the exit time may rise dramatically as a result. One avenue for future development is to bound the exit time by better incorporating it into the Bayesian model through a prior distribution that would limit the space of possible functions. Designing such a prior using Gaussian Processes is not trivial, however, as it needs to verify three boundary conditions imposed by the definition of the exit time (Lange-Hegermann, 2021) (see *Methods*).

## 4 DISCUSSION

Observed population frequencies have been commonly used to characterise alternative states in ecological communities and other complex systems (Costea et al., 2018; Livina et al., 2010). Recent studies used this approach, for instance, to identify bistability in human gut microbiota (Lahti et al., 2014) and climatology (Livina et al., 2010) in the absence of time series data. Despite the recent popularity of this approach, analyses based on times series can provide a more trustworthy and complete view of the stability landscape, indicating how, depending on the community state, the community fluctuates around the stable state or tends to drift towards more stable configurations of community composition. However, quantifying the stability landscape with scarce time series data remains challenging in many real applications.

We introduced a flexible method for reconstructing the stability landscape from limited observations. Our model quantifies changes in the drift and diffusion along the landscape, which permits the description of essential dynamical properties. Validation with real and simulated data demonstrate that the model can reliably locate stable equilibria and tipping regions, detect multistability using three orders of magnitude fewer points than other recent approaches (Arani et al., 2021), and estimate exit times. Related versatile models have been used to model thermal fluctuations in small particles in fluids (Atzberger, 2011), consumer behaviour (Krajbich et al., 2012), stocks (Braumann, 2019), and decision outcomes in neuroscience (Feltgen and Daunizeau, 2021). Whereas Gaussian Process regularisation has been leveraged to model financial and climatic data (e.g. (Garcia et al., 2017)), to our knowledge such models have not been developed yet in microbial ecology.

While our work demonstrates the utility of the probabilistic approach and joint analysis of multiple time series, one of the key limitations is the assumption that all observed short time series follow similar underlying dynamics; this limiting assumption could be relaxed by hierarchical extensions that incorporate individual deviations. Moreover, our model assumes that the overall dynamics does not change over time. Thus, changes in the stability landscape (Van Nes et al., 2016) or higher-order effects such as ecological memory (Khalighi et al., 2022) might set additional challenges. Time-varying models, or models where the landscape shape is governed by a covariate, could be a way forward. However, their added computational complexity is a key limitation for Gaussian Processes in large datasets with many time points. Finally, our results show that the inference of derived quantities such as the exit time may be particularly sensitive in regions of the state space where data is scarce.

This work addresses a timely challenge in areas such as human microbiota research, where long and dense time series are not commonly available and the system presents a mixture of overlapping and often poorly described processes. It also provides new insights into the modelling of complex ecosystem dynamics. In particular, the probabilistic approach extends the point-wise concepts of stable and tipping points to broader stable and tipping regions. This can naturally accommodate the uncertainty and fluctuations that are characteristic of many real systems (Arani et al., 2021). Our non-parametric approach does not require information on the underlying mechanisms, and is as such potentially applicable to a broad range of complex systems beyond ecological applications.

# ACKNOWLEDGEMENTS

We would also like to thank Egbert van Nes for discussions and suggestions. We acknowledge Raziye Rabiei for her contributions to the design of the first figure. Finally, we wish to acknowledge CSC – IT Center for Science, Finland, for generous computational resources. This project was supported by the Research Council of Finland (decision 330887 to L.L. and 340314 to G.S.K.), the Strategic Research Council in Finland (decision 352604 to L.L.), the University of Turku Graduate School (C.R.) the European Union's Horizon 2020 research and innovation programme (grant agreement No 952914 to L.L.), and the Sakari Alhopuro foundation (decision 20210172 to G.S.K.).

# CONFLICT OF INTEREST STATEMENT

The authors declare no competing interests.

# AUTHOR CONTRIBUTIONS

L.L. and V.L. conceived and initiated the study. W.d.V. provided the gut microbiota profiling data. C.R. planned and conducted the computational experiments, implemented the code, analysed the data, generated the figures, and wrote the first version of the manuscript. G.S.K. and L.L. supervised the project. C.R., G.S.K., and L.L. wrote the manuscript. All authors provided feedback during the work, read and approved the final version.

# DATA AVAILABILITY STATEMENT

The data and source code used in the analyses is openly available on GitHub: https://github.com/exod1a/gp_project.
The code was written in R version 4.1.0 and rstan_2.21.2 (Stan Development Team, 2024).

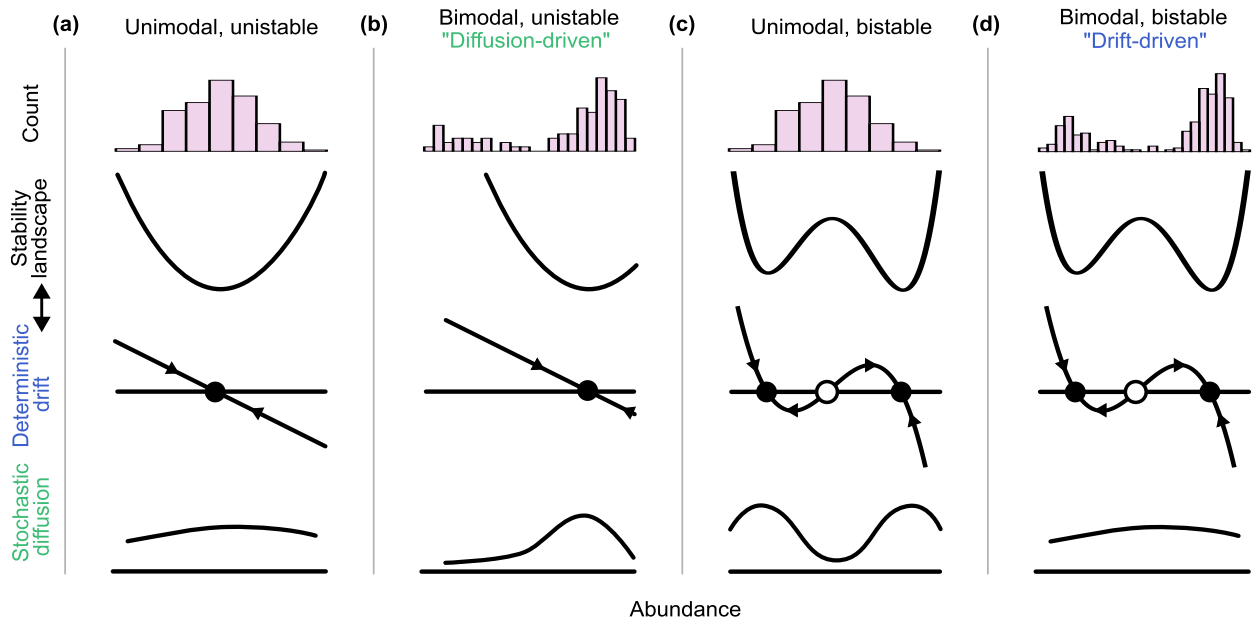

**Figure S1** An alternative representation of the distinction between modality and stability. The drift function quantifies the stability landscape but diffusion plays a key role in the type of data we observe. (a) Unistable drift with approximately constant diffusion can generate unimodal stationary densities. (b) Unistable drift and non-constant diffusion can generate "diffusion-driven" bimodal stationary densities. (c) Bistable drift with highly variable diffusion can generate unimodal stationary densities. (d) Bistable drift, combined with approximately constant diffusion generates "drift-driven" bimodal stationary densities.

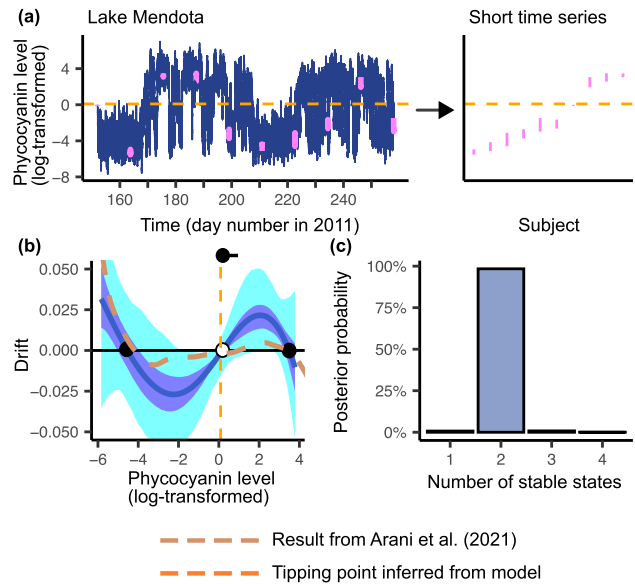

**Figure S2** Re-analysis of Lake Mendota data with fewer time points. (a) A recent study on Lake Mendota (Arani et al., 2021) used 60,000 equally spaced time points (left). We selected a small subset of 50 time points evenly distributed over the data range to simulate collection of short and sparse time series (right); the vertical pink lines represent these 10 short time series of 5 points. The dashed orange line the tipping point inferred by our model. (b) With only 50 time points, the drift function result from Arani et el. (2021) (dashed brown line) fell within the 95% credible interval of our model using three orders of magnitude fewer time points ($5 \times 10^1$ compared to $6 \times 10^4$). Shown in the upper margin is the tipping region. (c) Corresponding multistability posterior. Our Bayesian formulation provides 98% support for bistability in this ecosystem, in line with previous results. This outlines the model's ability to accurately identify stability characteristics including the locations of the stable modes with very few time points. Nevertheless, the stability of the diffusion estimate greatly decreases (not shown).

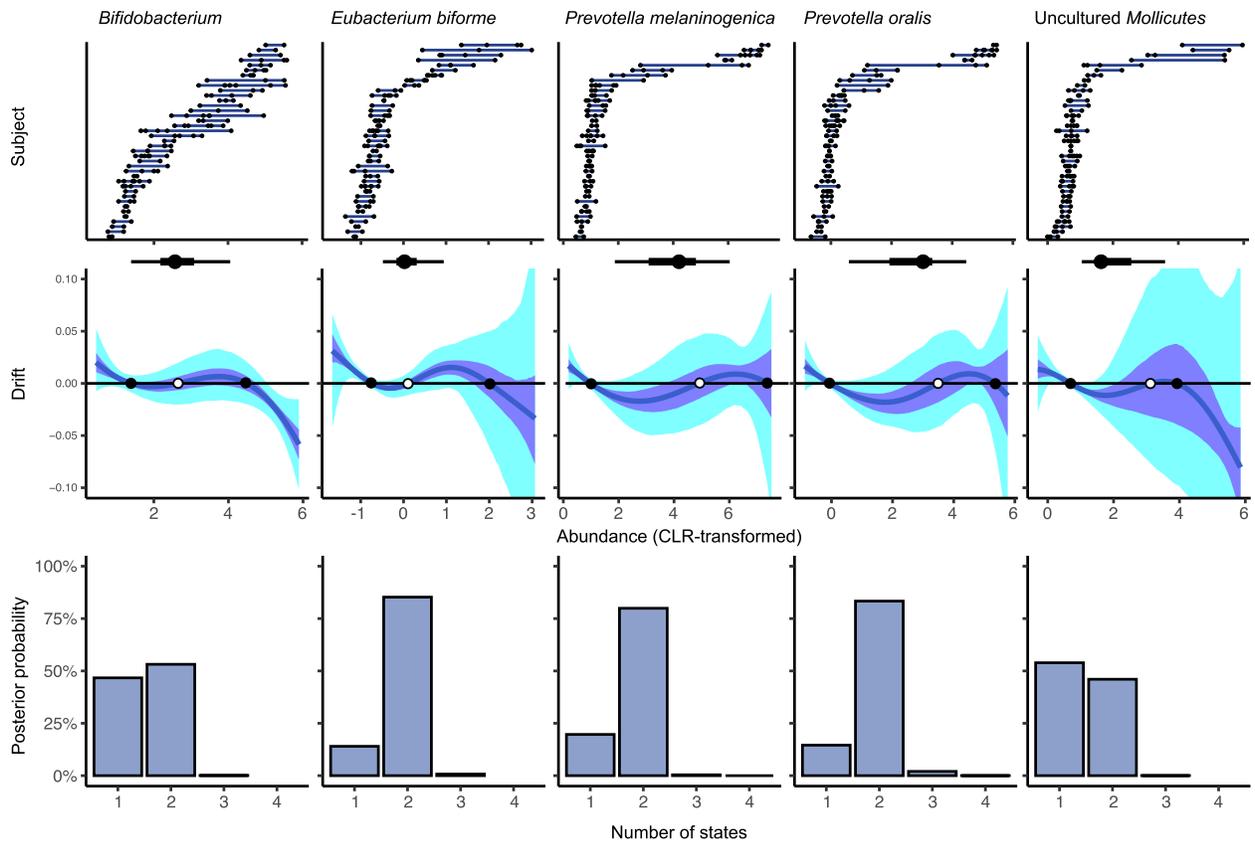

**Figure S3** Five taxa in the HITChip data set that show evidence of bistability. Each time series (upper row) consists of 2-6 time points for 39 subjects (total of 127 points). The posterior drift outputs with 50% and 95% credible intervals enveloping the means, including the stable modes (black points), tipping points (white circle) (middle row), and tipping regions in the top margins with their own means and 50% and 95% credible intervals (top whisker plots). Corresponding multistability posteriors (bottom row).

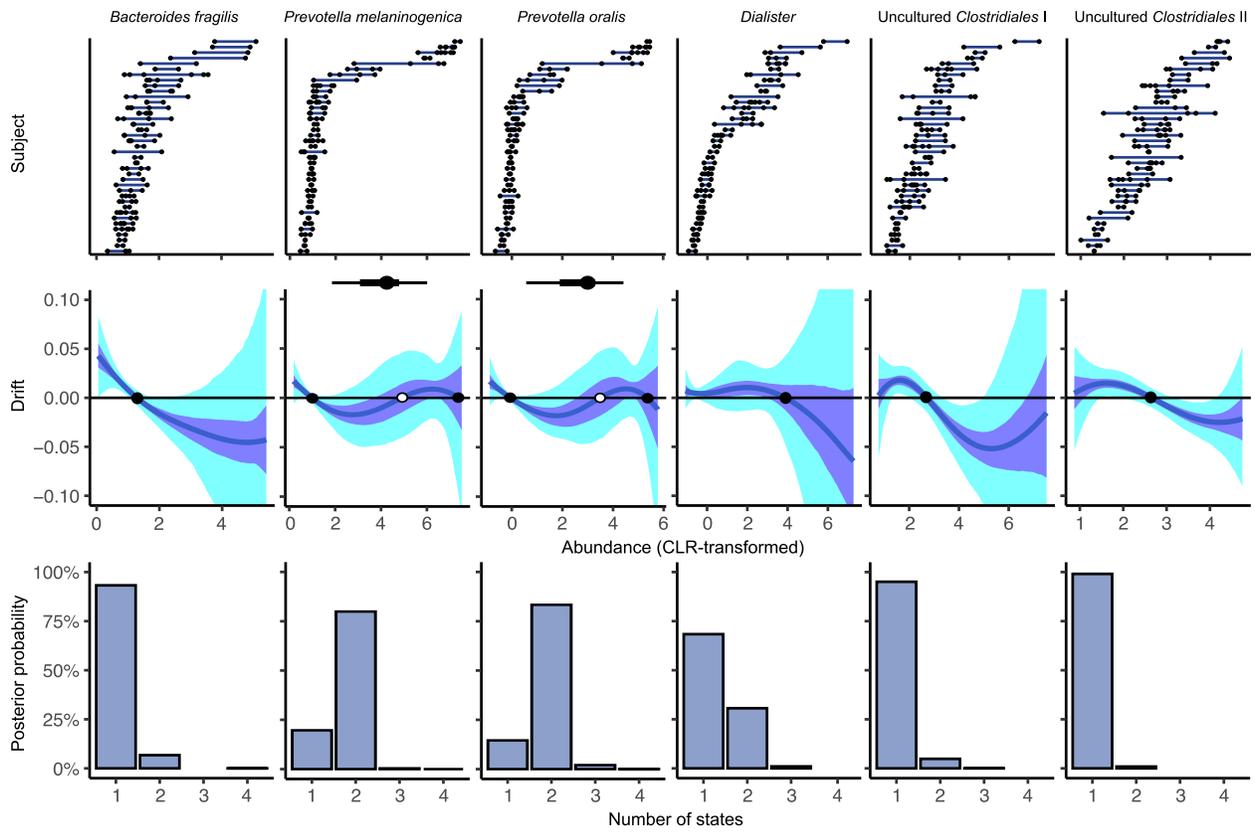

**Figure S4** The six taxa identified as tipping elements in Lahti et al. (2014). Each time series (upper row) consists of 2-6 time points for 39 subjects (total of 127 points). The posterior drift outputs with 50% and 95% credible intervals enveloping the means, including the stable modes (black points), tipping points (white circle) (middle row), and tipping regions in the top margins with their own means and 50% and 95% credible intervals. Corresponding multistability posteriors (bottom row).

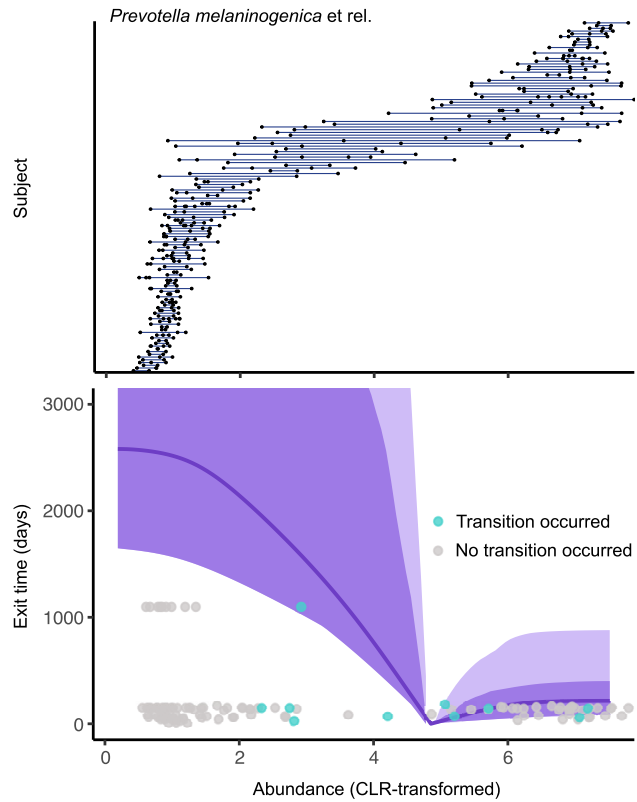

**Figure S5** Transition time scales present in the gut microbiota. The time series from the 128 healthy adult subjects (2-6 time points each, 358 in total) across all time steps in the HITChip data set for *Prevotella melaninogenica* et rel. (upper). A zoomed-in version of Fig. 5b. Grey points represent short time series that started at the given state value and did not transition within the given time and red points represent those that did. This shows that the exit time as a function of the abundance is on the same order of magnitude as the transition time scales obtained from directly from the data